\documentclass[reprint,amsmath,amssymb,aps,prl,showpacs]{revtex4-1}
\usepackage{epsfig}
\usepackage{amsmath}
\usepackage{amssymb}
\usepackage{graphicx}
\usepackage{dcolumn}
\usepackage{hyperref}
\usepackage{bm}
\usepackage{CJK}

\begin{document}

\preprint{APS/123-QED}
\begin{CJK*}{GB}{}
\CJKfamily{gbsn}
\title{Local transport measurements at mesoscopic length scales using scanning tunneling potentiometry}
\author{Weigang Wang(Íõκ¸Õ)}
\author{Ko Munakata}
\author{Michael Rozler}
 \altaffiliation[Now at ]{Rush University Medical Center, Chicago, IL 60612}
\author{Malcolm R. Beasley}
\email{beasley@stanford.edu}
\affiliation{Geballe Laboratory for Advanced Materials,\\
Stanford University, Stanford, CA 94305
}%
\date{\today}

\begin{abstract}
Under mesoscopic conditions, the transport potential on a thin film with current is theoretically expected to bear spatial variation due to quantum interference. Scanning tunneling potentiometry is the ideal tool to investigate such variation, by virtue of its high spatial resolution. We report in this {\it Letter} the first detailed measurement of transport potential under mesoscopic conditions. Epitaxial graphene at a temperature of 17K was chosen as the initial system for study because the characteristic transport length scales in this material are relatively large. Tip jumping artifacts are a major possible contribution to systematic errors; and we mitigate such problems by using custom-made slender and sharp tips manufactured by focussed ion beam. In our data, we observe residual resistivity dipoles associated with topoographical defects, and local peaks and dips in the potential that are not associated with topographical defects.
\end{abstract}


\maketitle
\end{CJK*}

Mesoscoic transport, i.e., transport on length scales where the quantum nature of transport directly manifests itself, is of great interest in understanding the emergence of the macroscopic transport properties of a material from their microscopic origins.   Perhaps the best-known example is weak localization\cite{Altshuler1980,Bergmann1984,Berger2004}, where quantum interference in the local transport has profound implications on the macroscopic transport of a metal.  However, rarely are these mesoscopic processes in macroscopic samples observed directly.

The most direct studies involve nanostructures where, in essence, macroscopic leads are brought close to a small geometric constriction\cite{Altshuler1981,Washburn1986,Russo2008,Datta1990,Beenakker1991,Reimann2002,Guedon2012}.  Mesoscopic phenomena in macroscopic samples have been discussed theoretically \cite {Zyuzin1987,Spivak1991,Chu1990} but not observed experimentally due to the lack of a practical means of measuring transport at very short length scales.  In this {\it Letter}, we report the first observations of local mesoscopic transport on the nanometer length scale, using scanning tunneling potentiometry (STP). For reasons discussed below, we selected epitaxial graphene for this initial study.

The concept of STP is not new\cite{Muralt1986}, but its general application has heretofore been fraught with technical difficulties, the most problematic of which has been the phenomenon of tip jumping\cite{Pelz1990,Rozler2008}.  In tip jumping, under scanning conditions, the specific point of the STM tip where tunneling occurs  may jump abruptly from one position on the tip to another, leading to a concomitant jump in the measured potential.  Previously, relatively clean STP studies have only been possible using specific topographical defects of known character on very smooth surfaces\cite{Briner1996, Ji2012}.  The study of mesoscopic transport under more general conditions has been lacking due to the tip jumping problem.  We have found that tip jumping can be greatly mitigated through the use of very slender STM tips with small tip areas\cite{Wang2013}.

Graphene\cite{Geim2007,Berger2004} is of great interest for its unique electronic structure and transport properties (e.g., Klein tunneling).  Here, however, we are using epitaxial graphene, for which the Fermi level is well removed from the Dirac point. Consequently,  the transport is to a first approximation more appropriately thought of as that of a low carrier density metal  with relatively clean and flat surfaces.  Because of the low carrier density,  the various length scales relevant for mesoscopic transport are quite large.

It is well established in the literature that epitaxial graphene exhibits weak localization phenomenon at low temperature\cite{Berger2004,deHeer2010}. Magnetoresistance measurement confirms that our sample is in weak localization regime at the low temperature (17$K$) at which our measurements were carried out.The estimated transport length scales are summarized in Table \ref{graphenenumbers}, in descending order.  Under these conditions,  the charge carriers scatter multiple times before they are inelastically scattered ($l_{in}$) or phase mismatch occurs ($l_T$), and the transport is clearly mesoscopic.

As we shall show, there is considerable structure in the transport potential of epitaxial graphene at these length scales, not all of which is readily understood.  Hence our work not only demonstrates the general applicability of STP for mesoscopic transport studies.  It also raises some new questions in mesosopic transport, at least  as they present themselves in epitaxial graphene.

\begin{table}
\begin{center}
\caption{Transport length scales of the epitaxial graphene sample measured in this {\it Letter} at 17$K$, estimated from Hall measurement and magnetoresistance measurement.}
\begin{tabular}{rc}
 \\
Inelastic mean free path $l_{in}$ (17K) &  \quad60nm \\
Thermal length $l_T\equiv\sqrt{\hbar D/k_BT}$ (17K) & \quad40nm
\\
Elastic mean free path $l$ & \quad9nm  \\
Fermi wavelength $\lambda_F$ & \quad4nm \\
\end{tabular}\label{graphenenumbers}
\end{center}
\end{table}

STP\cite{Muralt1986} uses a scanning tunneling microscope (STM) to characterize topography and to measure the spatial dependence of the local transport potential with high spatial resolution. The instrument used for this work is described in \cite{Rozler2008}. We note that due to the measurement protocol employed, %at each position of the STM tip, the topography and potential measurements are taken successively.
%For topography a dc bias voltage is applied to the STM tip, whereas for potentiometry, this tip  voltage is set to zero, and the transport potential is measured using lock-in detection in the presence of an ac bias sample transport current.
%The procedure ensures that
the dc bias voltage for STM imaging is turned off during potential measurement, and sensitivity of the measurement to vibrations is eliminated. Thus, only a minimal vibration isolation is necessary to achieve low noise and high spatial resolution. The STP measurement is a low frequency ac measurement.

\begin{figure}
\begin{center}
\includegraphics[width=2.2in,bb=0 0 1363.5  891]{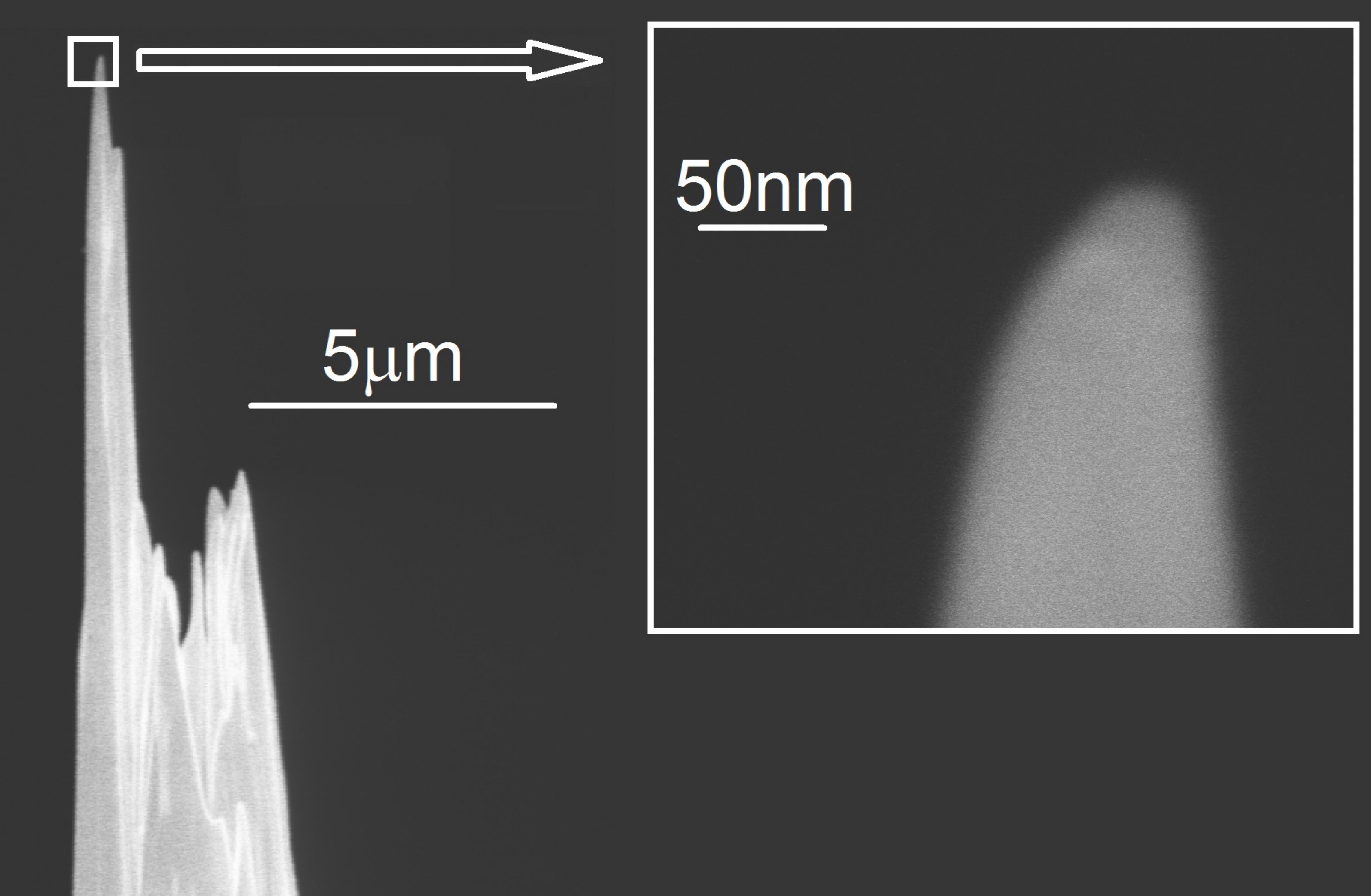}
\caption{SEM image of sharp STM tip used in the STP measurement, manufactured by FIB milling. Data in Figure \ref{STMSTP} were taken using this tip. From the flat surface topography of the sample and the shape of the tip, we set an upperbound of 30nm for possible tip jumping distance.}\label{tip}
\end{center}
\end{figure}

In order to mitigate the problem of tip jumping, we have used slender, sharp tips  micro-machined using a focussed ion beam (FIB)\cite{Vasile1991}. Progress during manufacture is monitored using scanning electron microscopy (SEM). Figure \ref{tip} shows an SEM image of a sharp platinum/iradium tip that was used in this work.  It was necessary to take SEM images both before and after use, to assure that the sharpness of the tip was not compromised during scanning.   As expected, large, abrupt tip jumping artifacts are absent in our data. While tip jumping can possibly account for small features in STP, most potential features in our data are larger than what can be expected from tip jumping.

Our samples were grown using the well-known approach\cite{Berger2004} of thermal processing the surface of a SiC (0001) substrate. In a vacuum better than 10$^{-8}$ torr, the substrate was heated to 830$^\circ$C, and then Si was deposited by pulsed laser deposition to desorb SiO from the surface. Following this, the sample was heated up slowly to 1480$^\circ$C and the surface was reduced to graphene.  Using Raman spectroscopy, the graphene thickness was estimated to be 3 to 4 layers. More characterization results can be found (as the control samples) in reference \cite{Lee2010}.

%Figure \ref{large}(b) shows the resistivity vs. temperature for this sample; the upturn of resistivity at lower temperatures at about 80K  indicates the onset of weak localization\cite{Berger2004}. In weak localization, the inelastic mean free path is larger than the elastic mean free path, which is consistent with our estimate in Table \ref{graphenenumbers}.

\begin{figure*}
\begin{center}
\includegraphics[width=5.5in,bb=0 0 1000.50 1431]{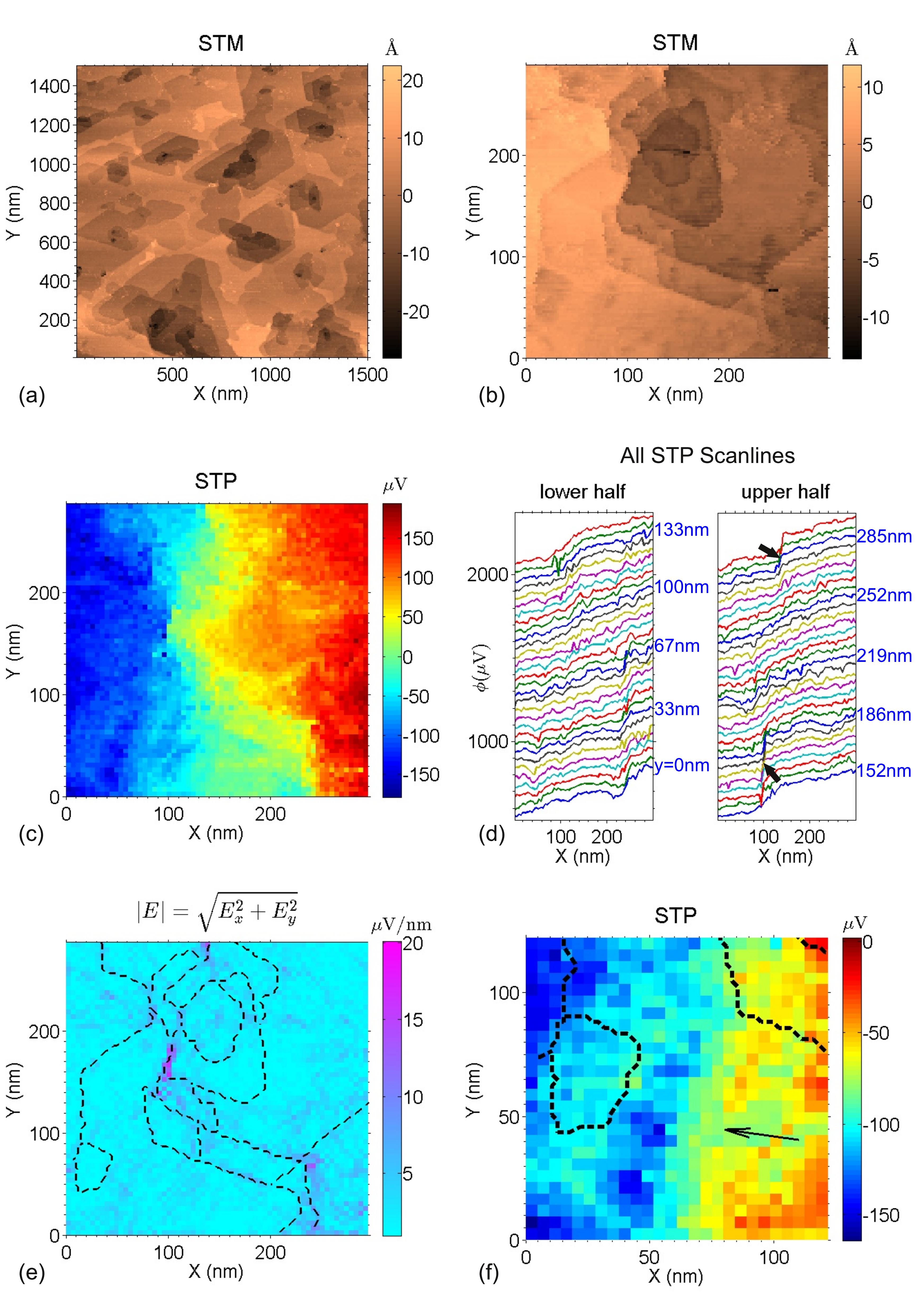}
\caption{Representative STM and STP data at 17K on epitaxial graphene. (a) Large scale STM topography at room temperature; (b) STM topography at 17K; (c) STP potential map of the same region in (b), the same set of data are plotted in (d);(d) All STP scan lines in (c), data from different scan lines are offset for clarity; (e) Magnitude of local electric field calculated from (c); (f) zoom-in of lower left region in potential map, showing peaks and dips in the potential, note that the color scale is different to more clearly show the features. The dashed lines in (e) and (f) are where edges are found in (b); the solid black lines in (f) are contour lines of $V=-130\mu V$; and the solid arrow in (f) represents the linear-fit electric field direction.}\label{STMSTP}
\end{center}
\end{figure*}

 Large-scale STM scans (see Figure \ref{STMSTP}(a)) show  that our samples are atomically flat and continuous, consisting of plateaus separated by well defined steps.  This is consistent with what others have observed \cite{Hass2008,Choi2010,deHeer2010,Hu2012}.  Typically, the  size of these plateaus is of the order 150nm. More detailed STM studies by others on similar samples have identified atomic scale defects as well as defects existing along plateau edge lines \cite{Rutter2007,Mallet2007,Lauffer2008}. These are presumably the disorder in the sample; and weak localization effects rise from carrier scattering by such disorder.%5230 edge points in 256 by 256 scan, scan size 1500nm, 1500/(5230/(2*256))=150nm

%\begin{figure}
%\begin{center}
%\includegraphics[width=1.5in,bb=0 0 399.83 408.08]{RT.pdf}
%\caption{Sheet resistivity versus temperature of one of the epitaxial graphene samples grown in our group. The minimum at about 80K indicates that the sample is in weak localization region below 80K.}\label{RT}
%\end{center}
%\end{figure}

\begin{figure}
\begin{center}
\includegraphics[width=3in,bb=0 0 2724  1184]{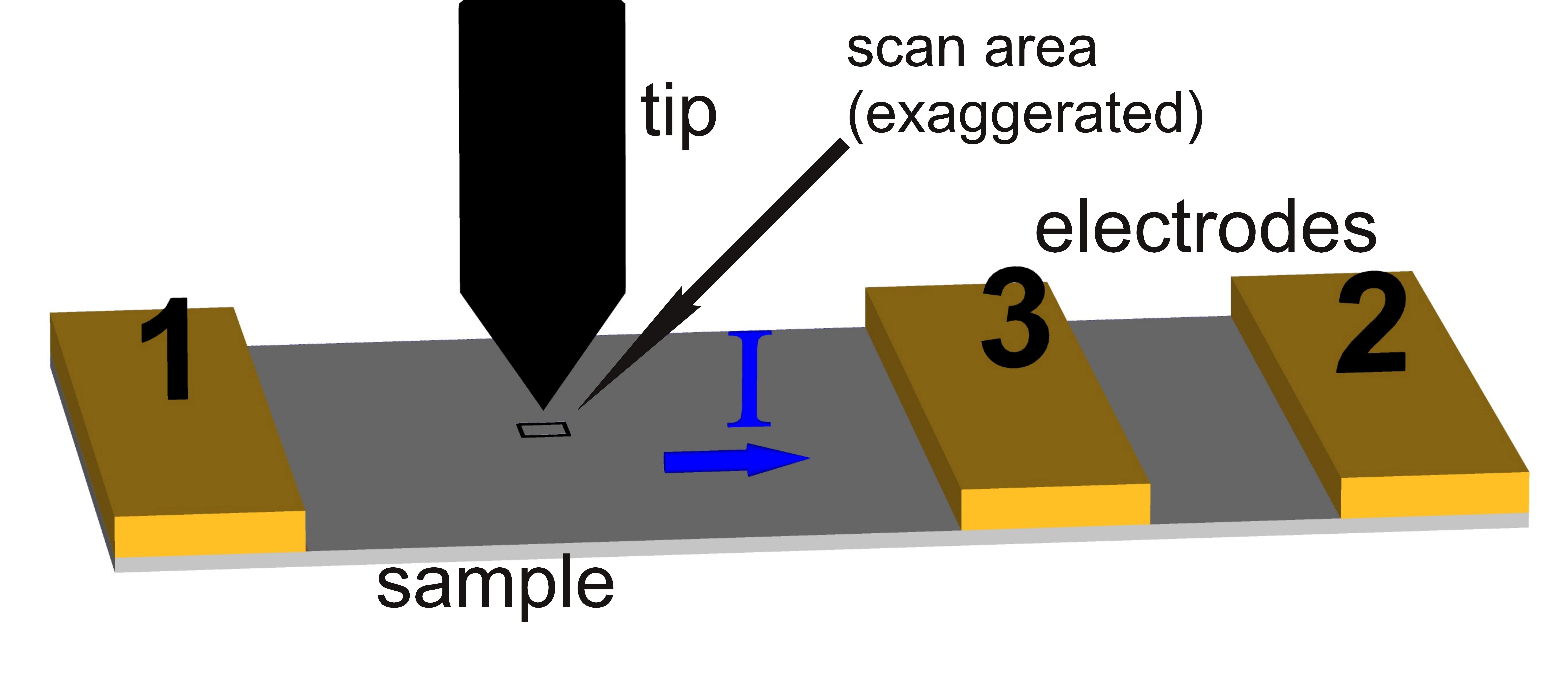}
\caption{Experimental scheme of STP measurement. The STM tip is brought to tunneling contact with a macroscopic sized sample. Planar, large electrodes are used to deliver the ac current. Potential across a much smaller area of the sample than the electrode geometry is sensed.}\label{scheme}
\end{center}
\end{figure}

The STP instrument introduced above was applied to the sample of size 5 mm, with large, planar electrodes to deliver the ac current (see Fig. \ref{scheme}). The scan size was much smaller than and far removed from the electrodes, and hence the effect of the electrode shapes can be neglected. The measurement was performed at 17K. In Figure \ref{STMSTP}(b) and (c), we present the STM and STP images on the same region.  The specific region shown was selected because it had strong features in the STP data.   As seen in Figure \ref{STMSTP}(c), the measured potential is high on the right and low on the left, reflecting the current flow in the sample. The average gradient of the potential is consistent with that deduced macroscopically from the electrode spacing and the voltage drop across the sample. If the sample were homogeneous, the local current density would be uniform and the contour lines of the potential map would be parallel straight lines; this is clearly not the case here.
%We now analyze the data in more detail, to show that while some of the features, mostly residual resistivity dipoles (RRDs), are dominant in the data and have been observed before\cite{Ji2012,Briner1996}, a fine structure even in the plateau area is present and cannot be understood in a classical diffusive picture.

%\begin{figure}
%\begin{center}
%\includegraphics[width=3.35in,bb=0 0 900.17 477.12]{largeandRT.pdf}
%\caption{Characterization of our epitaxial graphene sample. a) Large scale topography of epitaxial graphene at room temperature. b) Sheet resistivity versus temperature of one of the epitaxial graphene samples grown in our group. The minimum at about 80K indicates that the sample is in weak localization region below 80K.}\label{large}
%\end{center}
%\end{figure}

The most prominent features in the data are large potential jumps. We indicate two strong jumps (black arrows) in Fig. \ref{STMSTP}(d), in which all scan lines in Fig. \ref{STMSTP}(c) are plotted. These are Landauer redisual resistivity dipoles(RRDs)\cite{Landauer1957,Briner1996} centered at positions (110nm, 170nm) and (140nm,280nm). Note that  the exact shape of these dipoles depends on the geometry of the defect and on the nature of the transport in the vicinity of the dipole, and is different for, e.g., classical diffusive or fully quantum transport.  Given the characteristic length scales listed in Table \ref{graphenenumbers}, the transport should be mesoscopic over a distance of roughly 60nm around any static defect, well within the resolution of the data. These RRDs set the physical background of the fine structure that we will discuss shortly.

In order to more conveniently demonstrate the positions and strengths of the RRDs, in Figure \ref{STMSTP}(e), we show an image of the absolute value of the local electric field $|E|\equiv\sqrt{E_x^2+E_y^2}$ calculated from the potential data, on top of which is an overlay of the geometry of the edges seen in Figure \ref{STMSTP}(b).  As is evident in the figure, the strongest features in $|E|$ are associated with plateau edges.  At the same time, peculiarly, there are edges normal to the  current flow that show no features in the measured potential.

%Because it is in theory possible that the current is parallel to some edge lines locally due to the geometry of the defects, we carried out a two dimensional finite element simulation with the assumptions that: the conductivity on the plateaus are all the same; the conductivity at the edges are significantly smaller, and progressively smaller as the step height increases. We found that at edge positions where jumps in potential were not observed, there are potential jumps in the simulation. Thus the lacking of strong electric field regions in the data at these the edge line locations is not due to geometry.

The observation that most prominent features (pink in Figure \ref{STMSTP}(e)) are associated with plateau edges is consistent with the observation of Ji et al.\cite{Ji2012}.  However, our largest features are stronger than the edge features that Ji et al. measured on similar topographical structures. In addition, as noted above, there are edge lines with no RRD features.  Thus the observations in our sample suggest that the overall behavior in graphene is more complicated than is apparent in the work of Ji et al. The differences between STM topography and observed RRD locations in STP shows that while STM measures the topography of the sample surface, STP is an instrument sensitive to the internal properties of the defects.

%An interesting question regarding our data is whether it shows features specific to quantum transport or can be essentially understood classically.  Absent theoretical predictions, as a means of at least qualitatively addressing this question,  we carried out a simulation (on an assumed idealized geometry of our sample) of the potential image expected if the transport were purely classical diffusion.   To do this, we used  a two dimensional finite element simulation of classical diffusion on the geometry of our specific sample (dashed lines in Figure \ref{STMSTP} (b) and (c)). The results are shown in Figure \ref{STMSTP} (d) and (e). To capture the effects of edges, the simulation assumes a different conductivity over a distance of 5nm in the vicinity of the edges derived from Figure \ref{STMSTP} (a).  The conductivity assumed is $\sigma=(1+d/0.04\mbox{\AA})^{-1}\sigma_0$, where $d$ is the height difference across the edge, and $\sigma_0$ is the c\onductivity in the plateaus.  Thus the conductivity is progressively smaller as the step height increases.
%
%While the simulation clearly captures the gross features of the measured potential, there is fine structure that is not captured and presumably reflects the effects of quantum transport.  Morevoer,
%the simulation shows that the ``missing'' features in electric field in Figure \ref{STMSTP} (c) are not missing due to geometry.

There are other interesting features that we observed in the STP data. In Figure \ref{STMSTP}(f) we zoom in to the lower left region of Figure \ref{STMSTP}(c), with a smaller color scale range for better demonstration.  This region is of particular interest because it seems relatively free of the influence of large RRDs, and therefore should be more representative of homogeneous current flow.  The arrow in the figure shows the direction of the average electric field in this region. Evidently, features in this region are not trivial. First, the overall variation of the potential is in the form of peaks and valleys along the current flow; and is relatively constant perpendicular to the direction of the current flow.  Second, there are features in the form of peaks and dips in the potential with closed contour lines (e.g. the contour line enclosing (45nm, 20nm)). This cannot be understood in terms of classical diffusive transport in two dimensions as they would correspond to sources and sinks of current.

One might be tempted to attribute the closed contour lines  to electric fields associated with static charge trapped in graphene, as has been seen by other types of measurement, for example a scanning single electron transistor \cite{Martin2008}. This would be incorrect. The potential of static trapped charges is a dc eletrostatic potential, which is the patch effect. STP is not sensitive to electrostatic potential. To appreciate this point one can consider the case where there is no current. In this case the dc potential due to the static trapped charges is present, but measured STP would be zero everywhere on the sample, because when the sample is in equilibrium and the tip is at the same electrochemical potential, there is no tunneling current in the junction. Note that we do not dismiss the possibility that these STP features are due to the existence of trapped charges; however, a proper theoretical interpretation is needed to understand the reason why they did not, to first order, contribute a RRD, and to understand the mechanism for a trapped charge to contribute a peak or dip in the STP potential.

On these short length scales, the concept of electrochemical potential breaks down, and the physical interpretation of the measured potential is a subtle business  \cite{Wang2010}.  Still, the principle remains valid that one does not measure the electrostatic potential alone.

We also note that, in an STP measurement with a dc current, the temperature distribution on the sample and a temperature difference between the tip and the sample can lead to a spatial variation in the thermal voltage\cite{Druga2010}; in our measurement this effect is not present because we use an ac transport current for the STP measurement.

Finally, note that we do not observe ripply features in our images that have a periodicity  $\lambda_F$ , even near the strong RRDs.  This is in strong contrast to scanning tunneling spectroscopy (STS) images, where such strong ripples are often observed near defects \cite{Alpichshev2010,Wielen1996,Sprunger1997}. To understand this difference, we again emphasize that STS is sensitive to the electrostatic potential on the sample, which varies periodically due to the Friedel oscillation; whereas STP  measures the equivalent of the electrochemical potential and is not sensitive to this effect. On the other hand, at low temperature with an isolated defect, theoretical calculations have predicted weak Friedel-like oscillations\cite{Chu1990} in STP measurement, and our data sets an upper bound for such oscillations in epitaxial graphene at 17K.

In conclusion, the results reported here demonstrate that reliable direct measurement of transport on mesoscopic length scales is now possible. The data show the presence of Landauer RRDs associated with large steps in the topography of epitaxial graphene, but not at all such large steps. We have not examined thus far the internal structure of the RRDs that, in contrast to the diffusive transport case, should reveal fine structure due to quantum effects. In addition to the RRDs, we see structure in the transport potential in a relatively flat part of our STP image, this structure has no clear interpretation at this time.

%{\bf a brief introduction on STP theory}
%
%The peaks and dips we observe are not explicitly oscillatory as the Friedel-like oscillations above. As there is little theoretical attention predicting STP measurement result with model systems in a weakly localized sample, probably due to the lack of experimental data, we can only speculate at this stage. It is conceivable that at certain locations in the sample, the scattered carriers by different defects happen to interfere constructively in a way that gives an abnormal peak or dip, but since many defects are possibly involved, the locations of these peaks and dips are not explicitly oscillatory.

\begin{acknowledgements}
 Support for this work came from the Air Force Office of Scientific Research MURI Contract \# FA9550-09-1-0583-P00006. Two of us (WW and KM) further acknowledge the generous support of Stanford Graduate Fellowships. We would like to thank Nicholas Breznay for discussions on characterization of the weak localization in epitaxial graphene.
\end{acknowledgements}

%\bibliography{NewGrapheneBib}

%merlin.mbs apsrev4-1.bst 2010-07-25 4.21a (PWD, AO, DPC) hacked
%Control: key (0)
%Control: author (8) initials jnrlst
%Control: editor formatted (1) identically to author
%Control: production of article title (-1) disabled
%Control: page (0) single
%Control: year (1) truncated
%Control: production of eprint (0) enabled
%

\end{document}